\documentclass[twocolumn, prl, superscriptaddress]{revtex4}

\usepackage{graphicx}
\usepackage{amsmath}

\usepackage[colorlinks=true,linktocpage=true,linkcolor=blue,citecolor=blue,urlcolor=black]{hyperref}

\usepackage{lipsum}

\begin{document}

%\preprint{}

\title{
Centrality dependence of high energy jets in $p+$Pb collisions at the LHC 
}

\author{Adam Bzdak}
\email{bzdak@fis.agh.edu.pl}
\affiliation{AGH University of Science and Technology, Faculty of Physics and Applied Computer Science, PL-30-059 Krak\'ow, Poland}
\affiliation{RIKEN BNL Research Center, Brookhaven National Laboratory, Upton, NY 11973,
USA}

\author{Vladimir Skokov}
\email{vskokov@quark.phy.bnl.gov}
\affiliation{RIKEN BNL Research Center, Brookhaven National Laboratory, Upton, NY 11973,
USA}
\affiliation{Department of Physics, Western Michigan University, Kalamazoo, MI 49008, USA}

\author{Stefan Bathe}
\email{stefan.bathe@baruch.cuny.edu}
\affiliation{Baruch College, CUNY, New York, NY 10010, USA}
\affiliation{RIKEN BNL Research Center, Brookhaven National Laboratory, Upton, NY 11973,
USA}

\date{\today }


\begin{abstract}
The recently measured centrality dependence of high energy jets in proton-lead collisions at the LHC is investigated. We hypothesize that events with jets of very high energy (a few hundred GeV) are characterized by a suppressed number of soft particles, thus shifting these events into more peripheral bins. This naturally results in the suppression (enhancement) of the nuclear modification factor, $R_{pA}$, in central (peripheral) collisions. Our calculations suggest that a moderate suppression of the order of $20\%$, 
for $10^{3}$ GeV jets, 
can quantitatively reproduce the experimental data. 
We further extract the suppression factor as a function 
of jet energy and test our conjecture using available $R_{pA}$ data for various centralities.
\end{abstract}

\maketitle

%\section{Introduction}

Jet physics has proved to be very useful in uncovering properties of the hot medium created 
in heavy-ion (A+A) collisions \cite{d'Enterria:2009am}. A valuable baseline to study jet
quenching in A+A collisions is provided by proton-nucleus ($p+$A) collisions, where 
final state effects in the hot medium are expected to be suppressed. 
However, recent results on the centrality dependence of high energy 
jets in proton-lead ($p+$Pb) collisions at the LHC seem to challenge our understanding of
jet physics in nuclear reactions.   

To characterize the centrality dependence of jet production in $p+$A collisions and to 
compare to baseline proton-proton ($p+p$) collisions, one usually relies on the nuclear modification factor defined as  
\begin{equation}
R_{pA}=\frac{1}{\left\langle N_{\mathrm{coll}}\right\rangle }\frac{{dN_{%
\mathrm{jet}}^{pA}}/dp_{\perp }dy}{{dN_{\mathrm{jet}}^{pp}}/dp_{\perp }dy},
\label{Eq:RpA}
\end{equation}
where $\langle N_{\mathrm{coll}} \rangle$ is the number of nucleon-nucleon collisions and 
${dN_{\mathrm{jet}}}/dp_{\perp }dy$ is the average number of jets in $p+p$ or $p+$A. Both 
$\langle N_{\mathrm{coll}}\rangle $ and ${dN_{\mathrm{jet}}^{pA}}/dp_{\perp}dy$ are computed at a given centrality. 
$R_{pA}$ provides a quantitative value of the nuclear modification of the jet production rate 
relative to $p+p$ collisions, and deviations of $R_{pA}$ from unity
indicate non-trivial nuclear effects.
Following experimental collaborations, we will also consider the ratio of central-to-peripheral $R_{pA}$ defined as 
\begin{equation}
R_{cp} = \frac{\left. R_{pA} \right|_{\rm cent.}}{\left. R_{pA}\right|_{\rm periph.}}.
\label{Eq:Rcp}
\end{equation}

For jets of high transverse momenta  
$R_{pA}$ is expected to be close to unity based on perturbative QCD.  
A review of quantitative predictions is given in~Ref.~\cite{Albacete:2013ei}.
 
Recently ATLAS reported the dependence of the jet $R_{pA}$ on centrality, rapidity and 
transverse momentum~\cite{ATLAS:2014cpa}. Here we provide a brief overview of their findings. The measurements were performed for jets of very high $p_\perp$ and energies ranging from $40$ to roughly $2000$ GeV.  
The main points are: 
\begin{enumerate}
\item[a)] $R_{pA}$ is consistent with unity for minimum bias collisions (the centrality class $0-90\%$)
and does not demonstrate any systematic dependence on rapidity and transverse momentum. 
\item[b)] For proton-going rapidities ($y>0$), $R_{pA}<1$ in central collisions and 
$R_{pA}>1$ for peripheral ones. The effect increases as a function of
the jet transverse momentum and energy.
\item[c)] For backward rapidities (nucleus-going), $R_{pA}$ for $y<-0.8$ shows little dependence on 
transverse momentum and centrality, and is consistent with unity.
\item[d)] $R_{pA}$ of jets with $y>0$ approximately scales only with
  the total jet energy. 
\end{enumerate}

Recently the PHENIX Collaboration has studied  the bias from the
increased multiplicity of the underlying event when a hard trigger
particle is present~\cite{Adare:2013nff}.  
Its effect on $R_{pA}$ is 5\% or less in central events at RHIC and 20\% at LHC
energy.  However, it enhances $R_{pA}$ in central events, and
is thus opposite to the effect observed by ATLAS.

Alternatively, color fluctuations have been discussed as having an
influence on the correlation between a hard trigger and
the number of binary collisions in $p+$A
collisions~\cite{Alvioli:2013vk,Alvioli:2014sba}.  Also, it has been argued that
centrality estimators based on multiplicity measurements introduce a
bias on the hardness of the $p+$N collisions such that low multiplicity
$p+$A corresponds to lower than average number of hard scatterings~\cite{fortheALICE:2013xra} \footnote{After our paper was submitted to arXiv the effect of energy-momentum conservation on jet observables has been
studied in Ref.~\cite{Armesto:2015kwa}.}.

We will argue that the surprising observations of the ATLAS
experiment can be naturally understood assuming that events with jets of very high energy (a few hundred GeV) are characterized by a suppressed number of soft particles.  
We would like to illustrate this idea in
an extreme case.
Let us assume that events with high energy jets
are characterized by a strong suppression of soft particle production, 
so that the number of soft particles is of order $1$. 
Possible mechanisms of this suppression are of no relevance for this illustration 
and will be discussed at the end of this Letter. 
In the case of this strong suppression, all events with high energy jets will be counted 
as the most peripheral ones, by the usual procedure of centrality definition 
based on minimum bias events
(the most central events are defined as events with the largest number of soft particles). 
By construction, no jet events would fall into the most central class
and $R_{pA}$ for central collisions will be exactly zero. All events with high energy jets are counted as peripheral,
independently of the number of participants, 
 and in this case $R_{pA}>1$. It is also important to note that, in minimum bias
events,  the suppression of soft particle production does not 
influence $R_{pA}$ and consequently $R_{pA}=1$, unless there is another mechanism that modifies $R_{pA}$. 

In the following we discuss our model and present  quantitative results for $R_{pA}$. 
We argue that a moderate suppression of soft particle production of
the order of $20\%$, 
for the highest measured jet energies, allows to understand the data. 
Further we extract the suppression factor as a function of jet energy and test our hypothesis using 
available $R_{pA}$ data for various centralities.
 
We finish our Letter with
a discussion of possible mechanisms and conclusions.

\begin{figure}[t] 
\centerline{
\includegraphics[width=0.9\linewidth]{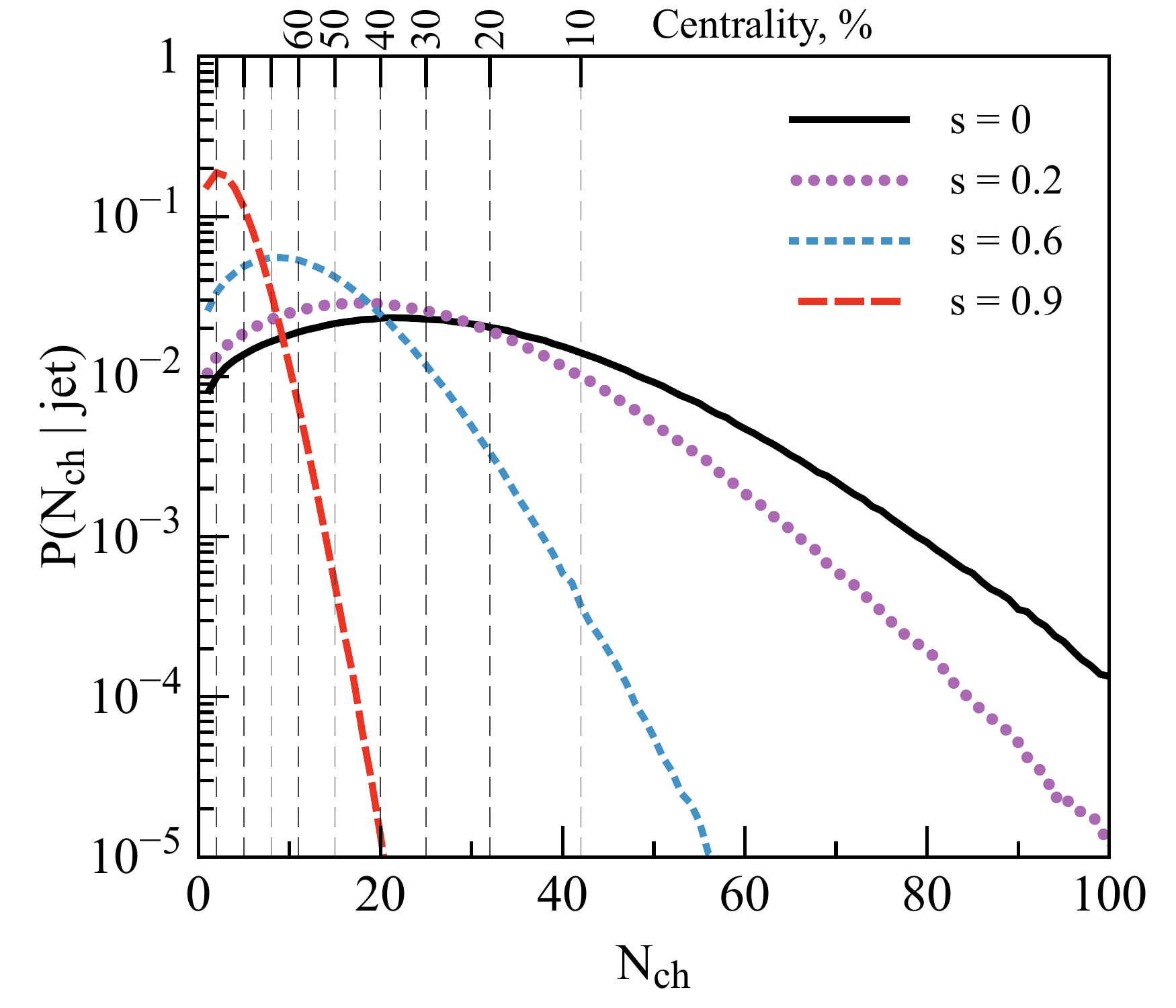} 
}
\caption{The probability distribution of the number of charged particles for events with high energy jets $P(N_{\rm ch}| \rm{jet})$ for the different suppression factors, $s$. The centrality classes are defined according to the  minimum bias probability distribution.  
} 
\label{fig:0} 
\end{figure}

\begin{figure*}[t] 
\centerline{
\includegraphics[width=0.45\linewidth]{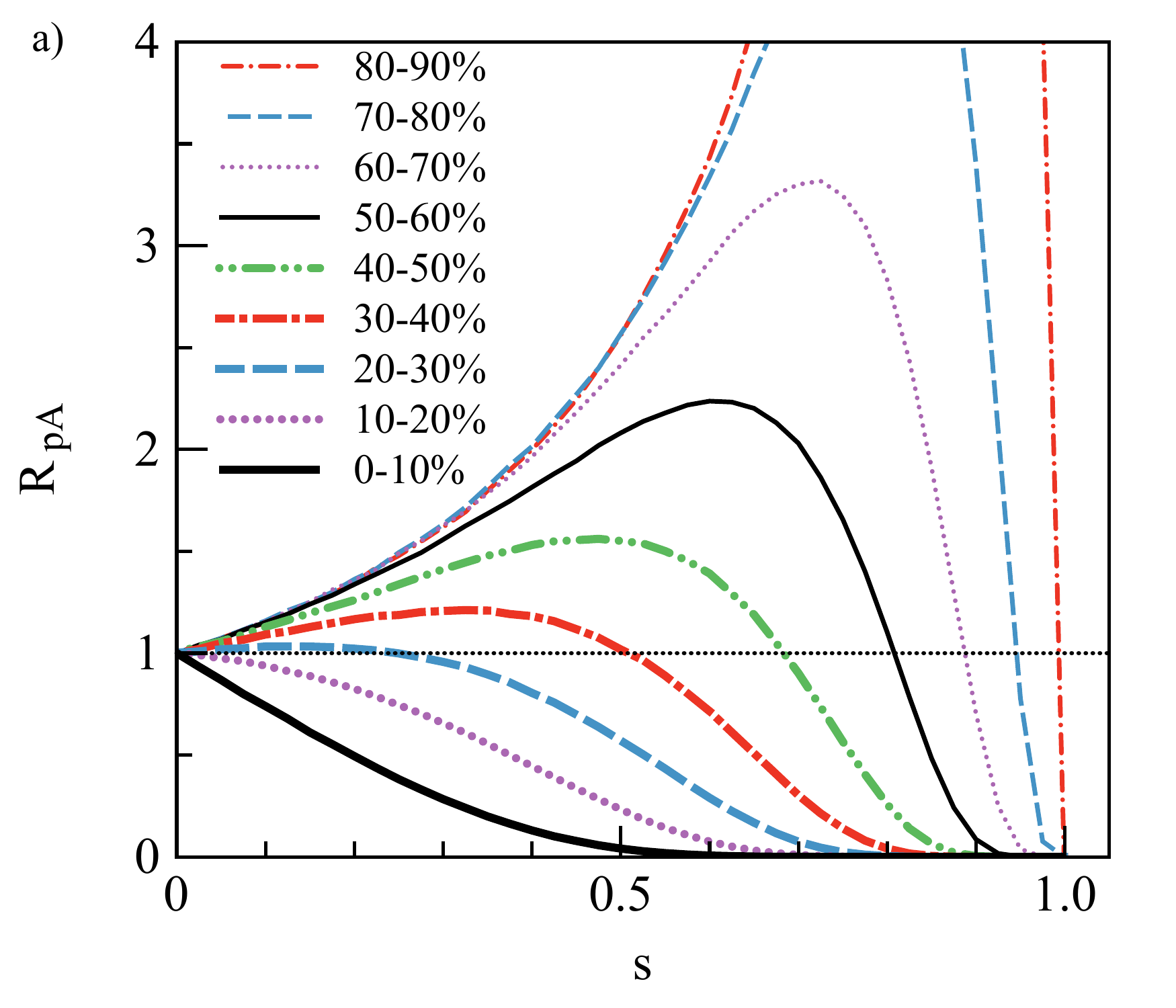} 
\includegraphics[width=0.45\linewidth]{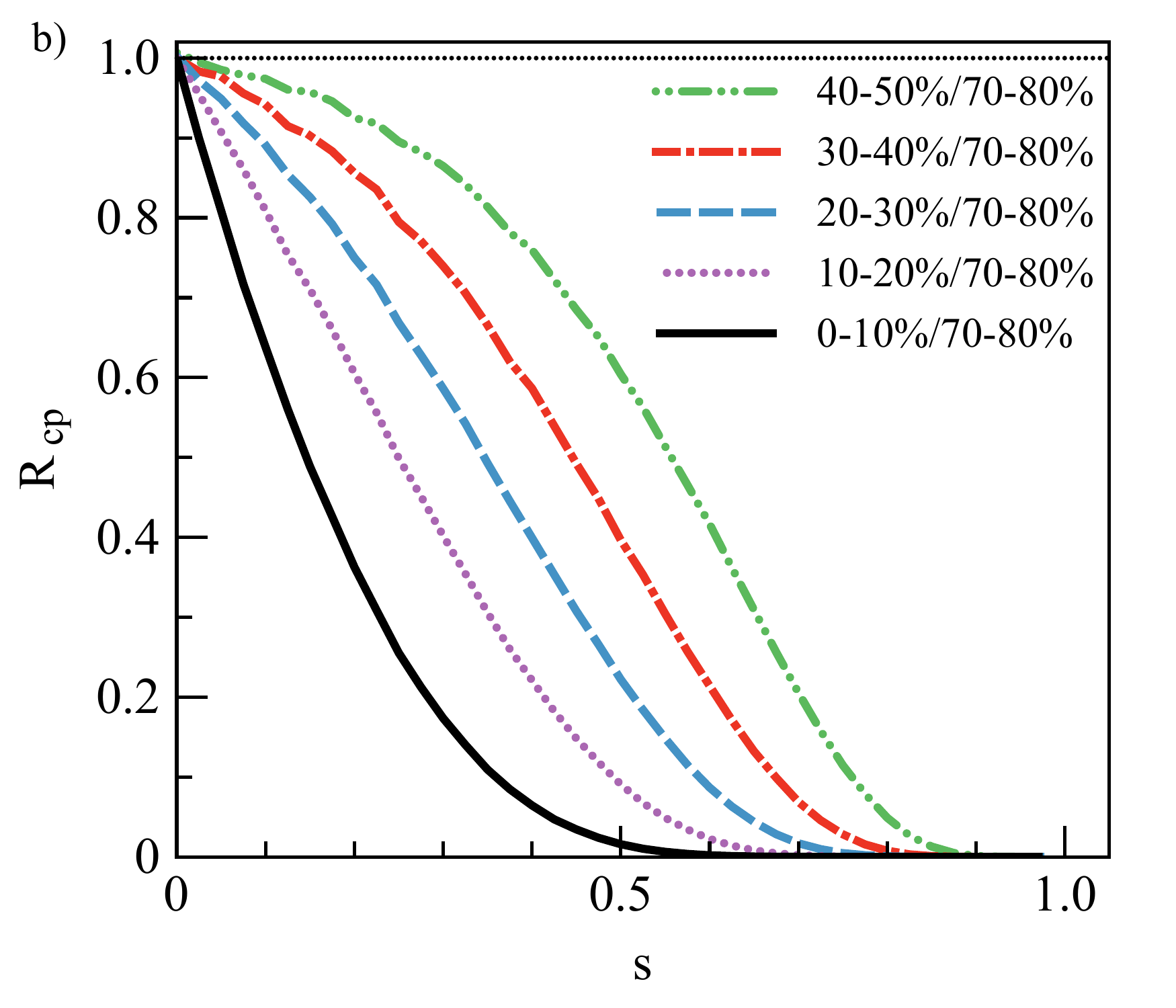} 
}
\caption{a) The nuclear
modification factor $R_{pA}$ in $p+$Pb collisions as a function of the soft particle
suppression factor $s$, see Eq. (\ref{Eq:s}), for various centrality classes defined by
the number of soft particles in minimum bias events. b) The central-to-peripheral ratio $R_{cp}$, Eq. (\ref{Eq:Rcp}), where the peripheral class is taken to be
$70-80\%$ centrality, as a function of the suppression factor $s$. In our model events with high energy jets are characterized by a suppressed number of soft particles thus shifting these events into more peripheral bins. This naturally results in the suppression (enhancement) of $R_{pA}$ in central (peripheral) collisions, respectively.
The suppression factor is expected to grow with jet energy or transverse momentum.
} 
\label{fig:1} 
\end{figure*}

%\section{Model}

\bigskip

In this Letter we assume the presence of anti-correlation 
between soft and very hard particle production, characterized 
by the suppression of soft particles in events with high energy jets.
We introduce the suppression in a general way, independent on the microscopic details of this effect, and 
demonstrate its effect on $R_{pA}$ and $R_{cp}$. The main points of our approach are listed below.

\begin{itemize}
\item Using the standard Glauber model~\cite{Miller:2007ri}, we evaluated the number
of inelastic nucleon-nucleon collisions, $N_{\rm coll}$, in each $p+$Pb event.
The distribution of nucleons inside a Pb-nucleus is given by the standard
Woods-Saxon distribution. The nucleon-nucleon inelastic cross-section is
taken to be $70$ mb~\cite{Zsigmond:2012vc}. 

\item We introduce the parameter $\epsilon\ll1$, the probability to produce
a high energy (or transverse momentum) jet in a {\it single} nucleon-nucleon
interaction.
It is evident that, in $p+$Pb events with large $N_{\rm coll}$,
the probability to produce a jet of high energy is larger. Mathematically, 
this probability is given by~\footnote{$1-\epsilon$ is the probability that no high-energy jet is produced in a single 
 nucleon-nucleon scattering. Thus $(1-\epsilon)^{N_{\rm coll}}$ is the probability that 
 no jet is produced in $N_{\rm coll}$ nucleon-nucleon scatterings.
 Consequently, Eq.~\eqref{Eq:ProbToJet} is the probability to produce at least one jet in $p+$A
 interactions with $N_{\rm coll}$ nucleon-nucleon scatterings.} 
\begin{equation}
1 - (1-\epsilon)^{N_{\rm coll}} \approx \epsilon N_{\rm coll},  
\label{Eq:ProbToJet}
\end{equation} 
where we assume that $\epsilon N_{\rm coll}\ll1$ to make our point clear.
Naturally, the value of $\epsilon$ depends on the jet energy, 
$E_{\rm jet}$:  the higher $E_{\rm jet}$ the smaller the value of $\epsilon$.
For high energy jets (so that the probability to produce more than one dijet is negligible) 
$\epsilon$ can be related to the jet yield according to
$\epsilon(p_{\perp},y) \propto \Delta p_{\perp} \Delta y dN_{\rm jet}/dydp_{\perp}$, 
where $\Delta p_{\perp}$ and $\Delta y$ are
narrow $p_{\perp}$ and $y$
bins around the measured $p_{\perp}$ and $y$. Consequently, the probability to produce a high energy jet is given by 
$\epsilon \propto \int dydp_{\perp} dN_{\rm jet}/dydp_{\perp}$, where the integration is over high values of $E_{\rm jet}$. In this Letter we are interested in jets with energies of order $1000$ GeV,
thus indeed $\epsilon$ is much smaller than one.

\item Now we determine the number of soft particles produced in a
$p+$Pb collision. In our model, the mean number of soft particles scales
linearly with the number of wounded nucleons~\cite{Bialas:1976ed,Bialas:2004su},
$N_{\rm part}=N_{\rm coll}+1$ \footnote{Possible deviations from this assumption, for instance 
originating from gluon saturation, see e.g. Ref.~\cite{Bzdak:2013zla},
does not change our conclusions.}. 
We further assume that each
participant populates soft particles according to a negative binomial
distribution (NBD),
which is known to approximate well measured multiplicity distributions in p+p interactions. The parameters of the NBD are chosen as follows: in an event
without a high energy jet the parameters for each participant are $\langle n_{pp}\rangle/2$ and $k_{pp}/2$ 
(so that in p+p we have $\langle n_{pp}\rangle$ and $k_{pp}$), where
$\langle n_{pp}\rangle$ and $k_{pp}$ are taken from fits to proton-proton collisions.
In our calculation we use $\langle n_{pp}\rangle = 5$ and $k_{pp} = 1.1$. In an event with a
high energy jet we assume that the mean number of soft particles from each participant is
reduced by a factor of $s$ as follows
\begin{equation} 
\langle n_{pp}\rangle\rightarrow \langle n_{pp}\rangle(1-s)
\label{Eq:s}
\end{equation} 
Clearly $0\leq s\leq 1$, ranging from no suppression to the
total suppression of soft particle production in events with high energy
jets. 
The suppression factor $s$ is a growing function of energy or transverse momentum of a jet.
The dependence of $s$ on jet energy will be discussed later.

\item Finally, we compute the centrality classes using the minimum bias multiplicity distribution with $s=0$ \footnote{$s>0$ does not change the centrality cuts since the probability to produce a high energy jet, 
$\epsilon$, is very small.} 
and for each centrality calculate $R_{pA}$ as a function of the suppression factor~$s$.
\end{itemize}

%\section{Results}

\begin{figure*}
\includegraphics[width=0.45\linewidth]{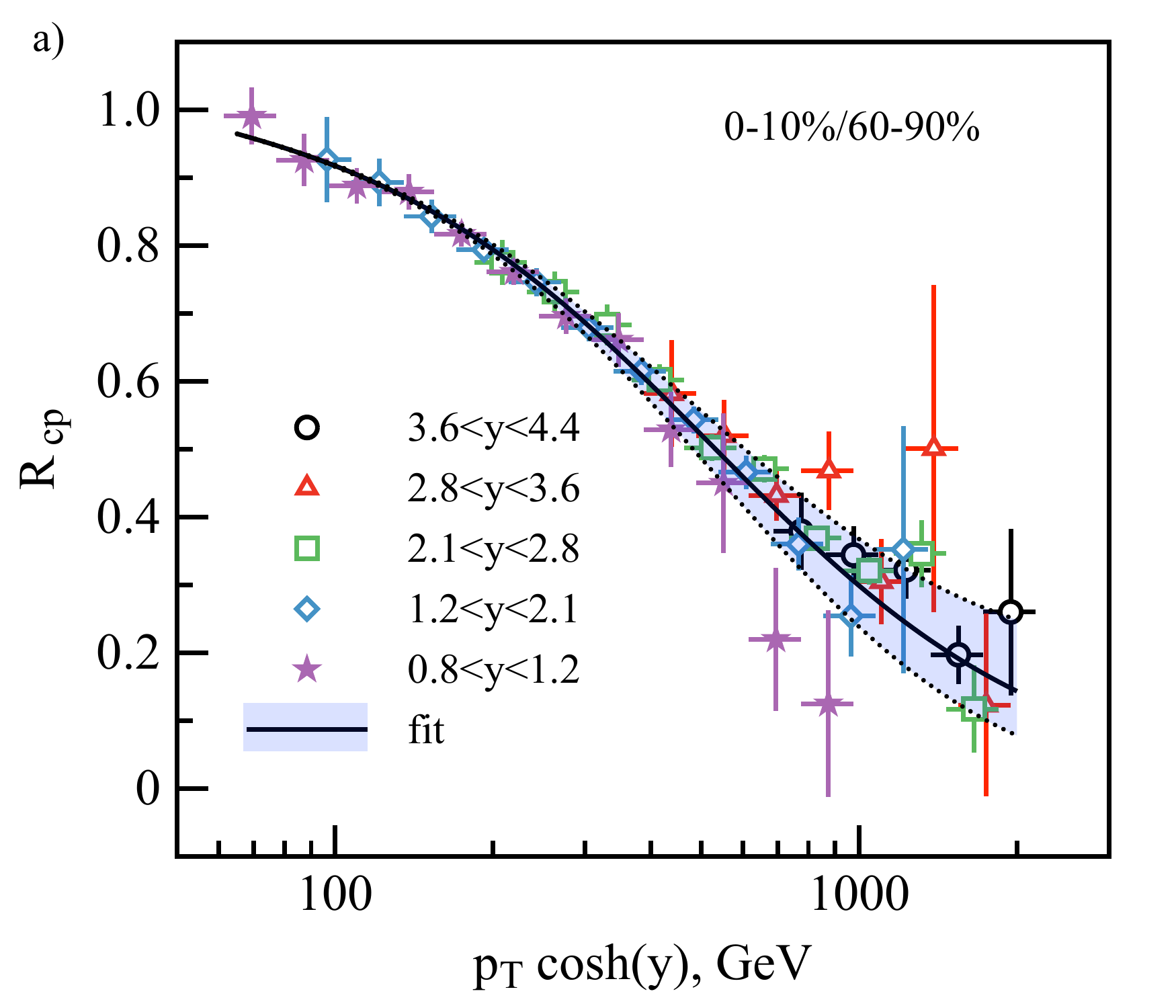}
\includegraphics[width=0.45\linewidth]{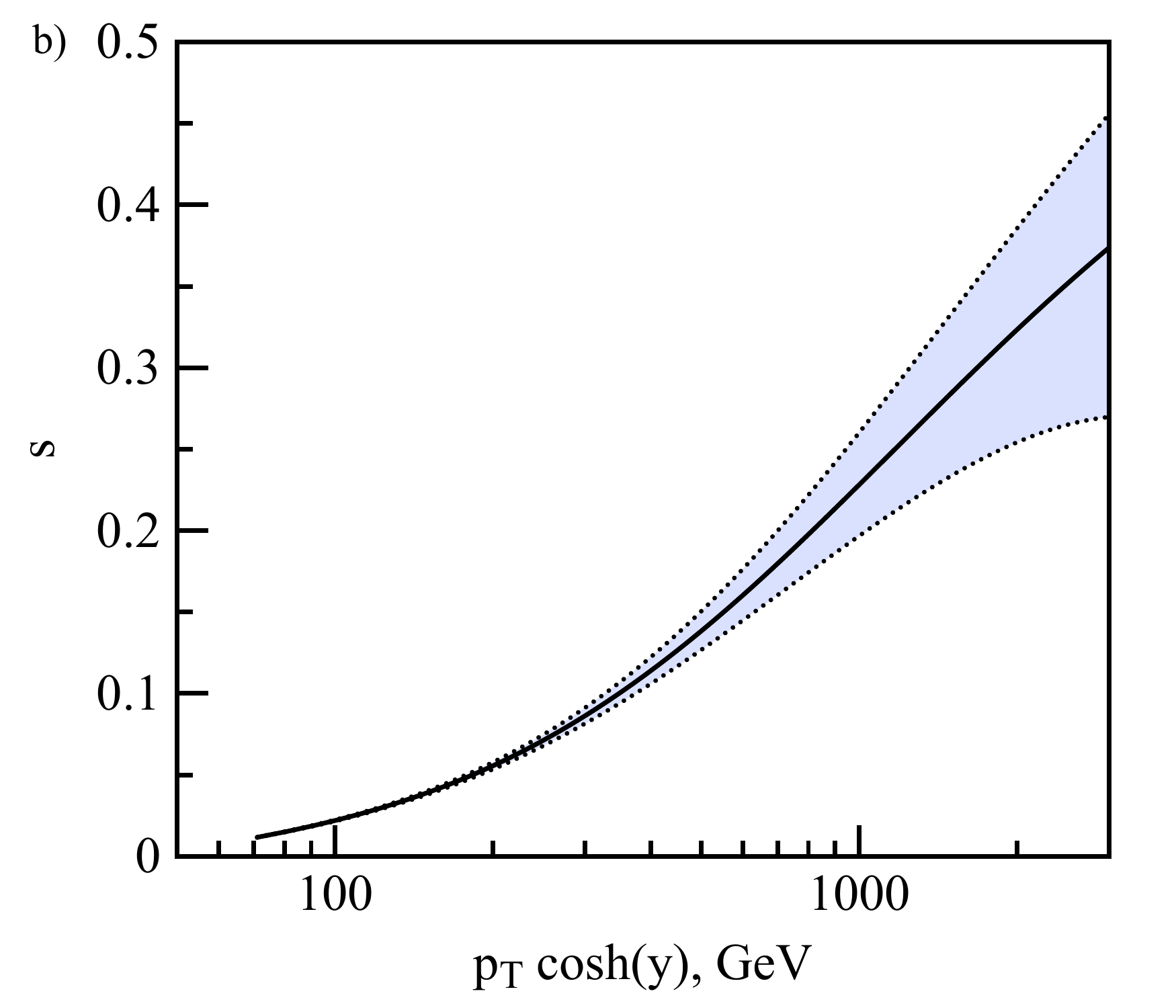}
\caption{a) The experimental data \cite{ATLAS:2014cpa} for $R_{cp}$ (the ratio of $0-10\%$ central to $60-90\%$ peripheral) as a function of $E_{\rm jet}=p_\perp \cosh(y)$ at different values of rapidity. 
The black line and the shaded area shows our fit with uncertainties. 
b) The corresponding suppression factor, $s$, dependence on $E_{\rm jet}$ extracted from the experimental data shown in the left plot.}
\label{fig:Rcpfit}
\end{figure*}

The results of our model calculation are presented in Figs.~\ref{fig:0} and \ref{fig:1} \footnote{It is worth noting that the results do not depend on the value of $\epsilon$,
provided $\epsilon N_{\rm coll}$ is much smaller than one. The only physics that can modify $R_{pA}$ is the postulated  suppression of soft particle production and thus $R_{pA}$ depends only on $s$.  
We explicitly checked that our results are practically unchanged for values of $\epsilon$ ranging from $10^{-3}$ to $10^{-5}$.}.

In Fig.~\ref{fig:0} we show the multiplicity distribution of charged particles produced in events with jets. The indicated centrality classes are computed from the minimum bias multiplicity distribution at $s=0$ and are unchanged throughout our analysis. The suppression $s>0$ modifies the multiplicity distributions by shifting them into more peripheral classes, as we discussed before.

This figure illustrates our mechanism. It is clear that jet events
with, say, $s=0.9$ are unlikely to be classified as central ones and
most likely will be classified as most peripheral. Thus $R_{pA}$
for the $0-10\%$ class is practically zero and is larger than $1$ for
the peripheral centrality class.

As illustrated in Fig. \ref{fig:1}, 
the case with no suppression results in   $R_{pA}=1$, because our model does not 
account for physics beyond soft particle suppression. 
For $s>0$ we observe the expected
enhancement for peripheral collisions and suppression for central events \footnote{For larger 
values of $s\to 1$ (strong suppression of soft particle production) we reach the limit where all events are
classified as peripheral and $R_{pA}\to 0$ for all bins, except the most
peripheral one.}.  

The ATLAS data~\cite{ATLAS:2014cpa} are in qualitative agreement
with the results from our Fig. \ref{fig:1}.

\begin{figure}[b]
\includegraphics[width=0.9\linewidth]{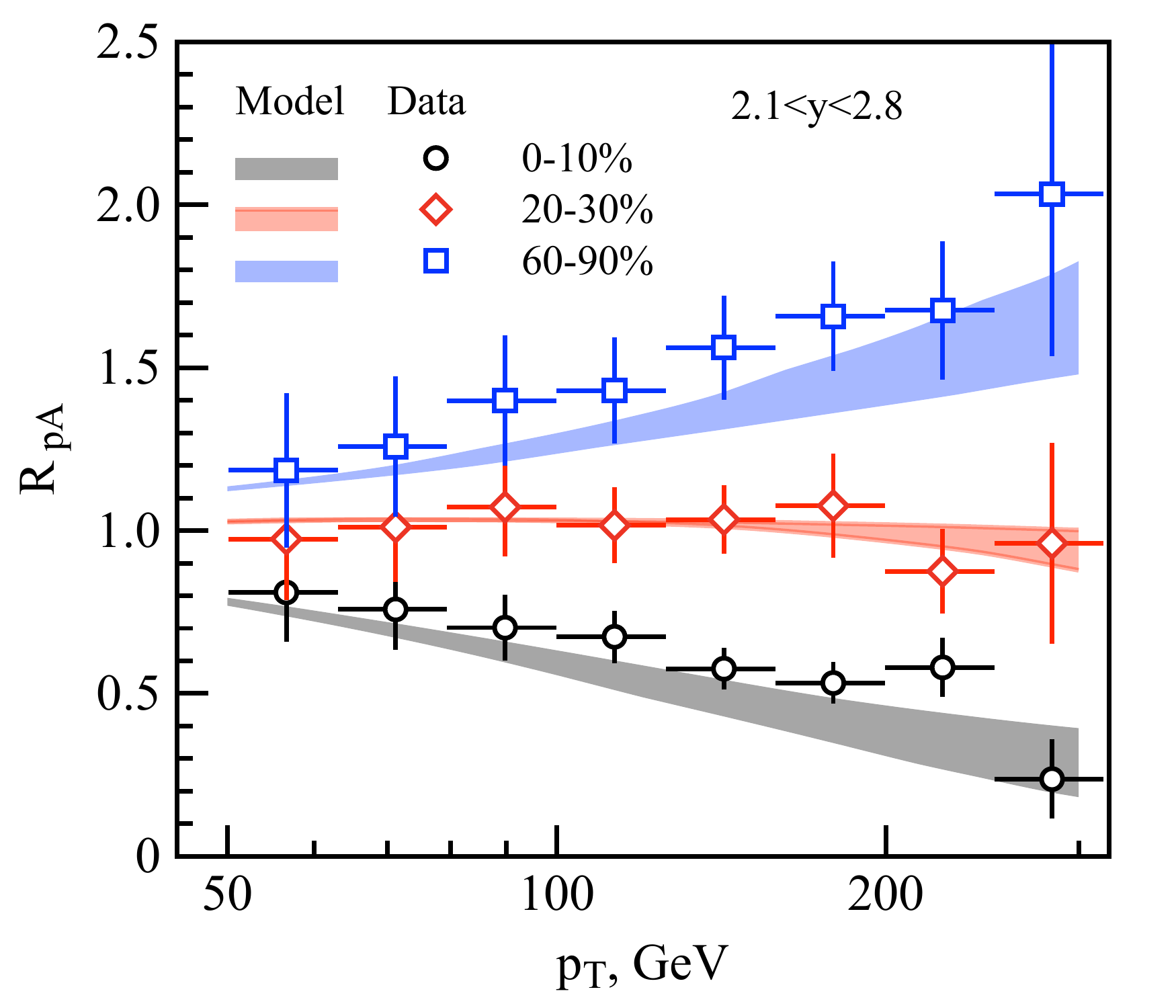}
\caption{The comparison of the model results (bands) with the experimental data \cite{ATLAS:2014cpa} (points) on $R_{pA}$ as a function of $p_\perp$ in the rapidity range $2.1<y<2.8$ for three centrality classes.}
\label{fig:RpaPred}
\end{figure}

To further check the model, we performed a fit of the available data for $R_{cp}$ 
(the ratio of $0-10\%$ to $60-90\%$) as a function of $E_{\rm jet} = p_\perp \cosh(y)$ for proton-going rapidities to extract the dependence of the suppression factor on $E_{\rm jet}$. This is illustrated in Fig.~\ref{fig:Rcpfit}. 
We conclude that the current data for jets of $1000$ GeV can be understood with
a moderately low suppression factor of the order of $0.2$.
Having extracted $s(E_{\rm jet})$ we can confront the model with the available experimental data on $R_{pA}$. In Fig. \ref{fig:RpaPred} we compare our model with $R_{pA}$ as a function of $p_\perp$ in the jet rapidity range $2.1<y<2.8$ for three different centrality classes. We observe satisfactory agreement \footnote{We extracted $s(E_{\rm jet})$ from the ratio of $0-10\%$ to $60-90\%$ $R_{pA}$ but it does not guarantee that the model works for separate $R_{pA}$ in these centrality classes.}.

%\section{Discussion}

To this point, we entertained the idea of suppression of soft particle production in events with 
high energy jets without providing a possible mechanism for this suppression. 
We want to stress that the problem at hand is highly nontrivial since it couples large (jets) and small (soft particles) $x$ physics, which is under poor theoretical control.  
Here, we will speculate about a possible mechanism that naturally explains
the forward-backward rapidity dependence of the ATLAS result. Let us consider the nuclear 
wave function as a function of $x$. Originally large $x$ partons evolve towards smaller $x$
by splitting (into smaller $x$ partons). During 
a collision the partons are liberated and eventually form final-state particles.
To produce a very large energy jet close to mid-rapidity,
the nuclear wave functions of 
both colliding objects should contain a large $x$ parton that 
did not suffer almost any splittings owing to a rare fluctuation in the evolution.
Since in a projectile proton one of the large $x$ partons is effectively removed 
from the evolution it cannot contribute to the production of small $x$ partons, and 
this results in the suppression of soft particle production. Thus the events with high energy jets effectively 
remove a large $x$ parton from a projectile proton, with the energy
proportional to the energy of the jet. Consequently for jets of very
high energy, we expect the reduction of soft particles to be roughly
$1/3$ \footnote{The number of produced particles scales with the
  product of two unintegrated gluon distributions, one of the
  projectile and the other of the target. By reducing the number of gluons in the projectile by $1/3$ we reduce the number of final particles roughly by the same factor.}, in agreement with our previous discussion.

The suggested mechanism of the suppression of soft particle production, 
in events with high energy jets, depends on the amount of energy removed from a projectile proton and thus should depend 
on the energy of a jet. This could explain the observed scaling of $R_{pA}$ and $R_{cp}$ with energy 
for different values of rapidity and transverse momentum.  
For a jet going into the forward direction (proton-going side) 
we expect the suppression to be stronger:  in order to produce such a
jet one needs to remove a large $x$ parton from a proton, whereas jets going into the nucleus direction
would require a large $x$ parton from a nucleus wave function.  The latter
does not activate the mechanism of suppression of soft particle
production \footnote{Removing a large $x$ parton form a nucleus can
easily be neglected since the number of participants is of the order of $10$.}. 
More detailed studies are underway.

%\section{Summary}

In conclusion, we propose a mechanism explaining the recently observed dependence of $R_{pA}$ and $R_{cp}$  
on centrality in $p+$A collisions. We show that a possible suppression
of soft particles in events with high energy jets naturally leads to
the observed suppression (enhancement) of $R_{pA}$ in central (peripheral) collisions, respectively. We found that a moderate soft particle suppression of the order of $20\%$ can provide a quantitative understanding of the ATLAS data. We compared the model with the data and found satisfactory agreement.

\vspace{-2mm}
\section{Acknowledgments}
\vspace{-4mm}

We thank P. Steinberg and L. McLerran for discussions.  
A.B. was supported by the Ministry of Science and Higher Education (MNiSW), by founding from the Foundation for Polish Science, and by the National Science Centre (Narodowe Centrum Nauki), Grant No. DEC-2014/15/B/ST2/00175 and in part by DEC-2013/09/B/ST2/00497.
The authors were supported through the RIKEN-BNL Research Center.
S.B. was supported by DOE award DE-SC0007017.

%%%%%%%%%%%%%%%%%%%%%%%%%%%%%%%%%%%%%%%%%%%%%%%%%%%%%%%%%%%%%%%%%%%


\begin{thebibliography}{99}

%\cite{d'Enterria:2009am}
\bibitem{d'Enterria:2009am} 
  D.~d'Enterria,
	  %``Jet quenching,''
		  arXiv:0902.2011 [nucl-ex].
			  %%CITATION = ARXIV:0902.2011;%%
				  %114 citations counted in INSPIRE as of 17 Jul 2014


%\cite{Albacete:2013ei}
\bibitem{Albacete:2013ei} 
  J.~L.~Albacete {\it et al.},
	  %``Predictions for $p+$Pb Collisions at sqrt s_NN = 5 TeV,''
		  Int.\ J.\ Mod.\ Phys.\ E {\bf 22}, 1330007 (2013)
			  [arXiv:1301.3395 [hep-ph]].
				  %%CITATION = ARXIV:1301.3395;%%
					  %55 citations counted in INSPIRE as of 17 Jul 2014
					  

%\cite{ATLAS:2014cpa}
\bibitem{ATLAS:2014cpa} 
  G.~Aad {\it et al.} [ATLAS Collaboration],
  %``Centrality and rapidity dependence of inclusive jet production in $\sqrt{s_\mathrm{NN}} = 5.02$ TeV proton-lead collisions with the ATLAS detector,''
  Phys.\ Lett.\ B {\bf 748}, 392 (2015)
  [arXiv:1412.4092 [hep-ex]].
  %%CITATION = ARXIV:1412.4092;%%
  %11 citations counted in INSPIRE as of 14 sept. 2015

%\cite{Adare:2013nff}
\bibitem{Adare:2013nff} 
  A.~Adare {\it et al.} [PHENIX Collaboration],
  %``Centrality categorization for R_{p(d)+A} in high-energy collisions,''
  Phys.\ Rev.\ C {\bf 90}, no. 3, 034902 (2014)
  [arXiv:1310.4793 [nucl-ex]].
  %%CITATION = ARXIV:1310.4793;%%
  %31 citations counted in INSPIRE as of 14 sept. 2015

%\cite{Alvioli:2013vk}
\bibitem{Alvioli:2013vk} 
  M.~Alvioli and M.~Strikman,
  %``Color fluctuation effects in proton-nucleus collisions,''
  Phys.\ Lett.\ B {\bf 722}, 347 (2013)
  [arXiv:1301.0728 [hep-ph]].
  %%CITATION = ARXIV:1301.0728;%%
  %15 citations counted in INSPIRE as of 12 Aug 2014

%\cite{Alvioli:2014sba}
\bibitem{Alvioli:2014sba} 
  M.~Alvioli, L.~Frankfurt, V.~Guzey and M.~Strikman,
  %``Revealing “flickering” of the interaction strength in pA collisions at the CERN LHC,''
  Phys.\ Rev.\ C {\bf 90}, 034914 (2014)
  [arXiv:1402.2868 [hep-ph]].
  %%CITATION = ARXIV:1402.2868;%%
  %9 citations counted in INSPIRE as of 14 sept. 2015

%\cite{fortheALICE:2013xra}
\bibitem{fortheALICE:2013xra} 
  A.~Morsch [ALICE Collaboration],
  %``p-Pb Results from ALICE with an Emphasis on Centrality Determination,''
  J.\ Phys.\ Conf.\ Ser.\  {\bf 509}, 012021 (2014)
  [arXiv:1309.5525 [nucl-ex]].
  %%CITATION = ARXIV:1309.5525;%%
  %3 citations counted in INSPIRE as of 12 Aug 2014

%\cite{Armesto:2015kwa}
\bibitem{Armesto:2015kwa} 
  N.~Armesto, D.~C.~Gülhan and J.~G.~Milhano,
  %``Kinematic bias on centrality selection of jet events in pPb collisions at the LHC,''
  Phys.\ Lett.\ B {\bf 747}, 441 (2015)
  [arXiv:1502.02986 [hep-ph]].
  %%CITATION = ARXIV:1502.02986;%%
  %3 citations counted in INSPIRE as of 14 sept. 2015

%\cite{Miller:2007ri}
\bibitem{Miller:2007ri}
  M.~L.~Miller, K.~Reygers, S.~J.~Sanders and P.~Steinberg,
  %``Glauber modeling in high energy nuclear collisions,''
  Ann.\ Rev.\ Nucl.\ Part.\ Sci.\  {\bf 57} (2007) 205
  [nucl-ex/0701025].
  %%CITATION = NUCL-EX/0701025;%%
  %485 citations counted in INSPIRE as of 13 Jul 2014

%\cite{Zsigmond:2012vc}
\bibitem{Zsigmond:2012vc} 
  A.~J.~Zsigmond [CMS Collaboration],
  %``Inelastic proton-proton cross section measurements in CMS at $\sqrt{s}=7$ TeV,''
  arXiv:1205.3142 [hep-ex].
  %%CITATION = ARXIV:1205.3142;%%
  %3 citations counted in INSPIRE as of 13 Jul 2014

%\cite{Bialas:1976ed}
\bibitem{Bialas:1976ed} 
  A.~Bialas, M.~Bleszynski and W.~Czyz,
	  %``Multiplicity Distributions in Nucleus-Nucleus Collisions at High-Energies,''
		  Nucl.\ Phys.\ B {\bf 111}, 461 (1976).
			  %%CITATION = NUPHA,B111,461;%%
				  %488 citations counted in INSPIRE as of 10 Jul 2014

%\cite{Bialas:2004su}
\bibitem{Bialas:2004su} 
  A.~Bialas and W.~Czyz,
	  %``Wounded nucleon model and Deuteron-Gold collisions at RHIC,''
		  Acta Phys.\ Polon.\ B {\bf 36}, 905 (2005)
			  [hep-ph/0410265].
				  %%CITATION = HEP-PH/0410265;%%
					  %66 citations counted in INSPIRE as of 10 Jul 2014
					  
%\cite{Bzdak:2013zla}
\bibitem{Bzdak:2013zla} 
  A.~Bzdak and V.~Skokov,
	%``Decisive test of color coherence in proton-nucleus collisions at the LHC,''
	Phys.\ Rev.\ Lett.\  {\bf 111}, 182301 (2013)
	[arXiv:1307.6168 [hep-ph]].
	%%CITATION = ARXIV:1307.6168;%%
	%6 citations counted in INSPIRE as of 10 Jul 2014

\end{thebibliography}
\end{document}